\documentclass[10pt,a4paper]{article}

\usepackage[top=0.85in,left=1.25in,footskip=0.75in,marginparwidth=2in]{geometry}
\usepackage{cite}
\usepackage{nameref,hyperref}
\usepackage{microtype}
\DisableLigatures[f]{encoding = *, family = * }
\usepackage{changepage}
\usepackage[aboveskip=3pt,labelfont=bf,labelsep=period,singlelinecheck=off]{caption}
\usepackage{lastpage,graphicx}
\usepackage{epstopdf}
\usepackage{graphicx}
\usepackage[T1]{fontenc}    
\usepackage{url}            
\usepackage{amsfonts}       
\usepackage{nicefrac}       
\usepackage{microtype}      
\usepackage{graphicx}
\usepackage{amsmath, amssymb}
\usepackage{bbm}

\title{Fast ABC with joint generative modelling and subset simulation}
\author{Eliane Maalouf \\ University of Neuchâtel \\ eliane.maalouf@unine.ch  \and
David Ginsbourger \\ University of Bern \\ david.ginsbourger@stat.unibe.ch \and
Niklas Linde \\ University of Lausanne \\ niklas.linde@unil.ch}
\date{}

\begin{document}
\maketitle
\begin{abstract}
We propose a novel approach for solving inverse-problems with high-dimensional inputs and an expensive forward mapping. It leverages joint deep generative modelling to transfer the original problem spaces to a lower dimensional latent space. By jointly modelling input and output variables and endowing the latent with a prior distribution, the fitted probabilistic model indirectly gives access to the approximate conditional distributions of interest. Since model error and observational noise with unknown distributions are common in practice, we resort to likelihood-free inference with Approximate Bayesian Computation (ABC). Our method calls on ABC by Subset Simulation to explore the regions of the latent space with dissimilarities between generated and observed outputs below prescribed thresholds. We diagnose the diversity of approximate posterior solutions by monitoring the probability content of these regions as a function of the threshold. We further analyze the curvature of the resulting diagnostic curve to propose an adequate ABC threshold. When applied to a cross-borehole tomography example from geophysics, our approach delivers promising performance without using prior knowledge of the forward nor of the noise distribution.   
\end{abstract}

\section{Introduction}
\label{sec:introduction}
Inverse problems encompass situations where unknown inputs are to be inferred based on given outputs like when inverting for physical parameters (e.g. geosciences, astrophysics, etc.) based on observations, or, in broad generality, when inferring parameters of statistical models relying on samples.     
Here, we focus on situations where a high-dimensional $X$ needs to be retrieved based on an observation of $Y=F(X)+ \eta$, for which calls to the ``forward model'' $F$ are costly and with $\eta$ being a noise term with an unknown distribution. Such inverse problems are generally ill-posed, rendering their solutions non-unique \cite{calvetti2018inverse}, requiring methods that recover the diversity of potential solutions.  
The Bayesian framework delivers a full posterior distribution for $X$ given $y$, a realization of $Y$.
It requires the specification of a prior distribution for $X$ and  a likelihood function, which relies both on $F$ and on the noise distribution. With this problem formulation, Markov Chain Monte Carlo (MCMC) algorithms are classically used to sample from posterior distributions \cite{robert2013MC}. 

Although MCMC approaches offer great flexibility, they still suffer from: the high cost of forward evaluations, a cost compounded by a high-dimensional $X$ requiring a large number of explored input instances; and the distribution of $\eta$ needs to be specified. 
MCMC methods were accelerated by reducing the number of calls to $F$ either by exploiting the geometry of the parameter space induced by the statistical model \cite[and references therein]{robert2018acceleratingMCMC}, or by pre-screening the most promising candidate inputs based on lower-fidelity approximations of $X$ as in two-stage MCMC \cite{2stageMCMC}. 
While lower-fidelity approximations can be obtained by simplifying the underlying model \cite[and references therein]{josset2015accelerating}, data-driven approximations relying on statistical and machine learning ideas have also been pursued \cite{gutmann2016bayesian,jarvenpaa2020parallel,jarvenpaa2017gaussian}. Considerable speed-ups can be obtained by surrogating $F$ but high input dimensionality and lack of knowledge of the noise distribution still pose notoriously hard problems. 

To circumvent noise specification, Approximate Bayesian Computation (ABC) is a set of likelihood-free methods that extensively sample from a prior distribution on the input space, running stochastic simulations emulating the forward model and the noise generating process, and accepting only candidate inputs that yield outputs close to the observed data \cite{mcmcwolikelihood}. Closeness is specified in terms of a dissimilarity measure on the output space and a tolerance level. The smaller the tolerance level is, the closer the posterior approximation gets to the true posterior but sample acceptance rates decrease \cite{sisson2018handbook}. 
Yet, when the input is high dimensional the cost of ABC is exacerbated by the increasing number of samples needed to efficiently cover the space and avoid missing relevant modes in the posterior. Adaptive methods for ABC aimed at improving its efficiency by sequentially tuning the proposal distribution in order to target promising regions in the input space \cite[and references therein]{robert2008adaptivity,chiachio2014approximate}. These methods still require calling the forward model during inference which can become a bottleneck.

Our proposed methodology addresses high dimensionality and costly forward models by leveraging joint Generative Neural Networks (jGNN) in combination with adaptive ABC principles. More specifically, a jGNN based on Sinkhorn Auto-Encoders (SAE) \cite{patriniSinkhornAE} parametrizes the candidate solutions by lower-dimensional latent vectors that are explored, offline, with ABC by Subset Simulation (ABC-SubSim) \cite{chiachio2014approximate}.

Our methodological contributions are: 
\begin{itemize}
    \item Development of an approximate inverse-problem solving framework relying on jGNN based on a generalization of SAEs to joint distribution modelling, 
    \item Efficient sampling of regions in the jGNN latent space susceptible to have generated the observed data using ABC-SubSim ``offline'', 
    \item Proposing a procedure to select the ABC threshold based on monitoring the approximate posterior distribution on the latent space through its estimated prior probability content as a function of the ABC tolerance threshold. 
\end{itemize}
The paper is structured as follows: Section \ref{sec:related_works} presents related works concerning deep generative modelling in probabilistic inference; Section \ref{sec:methodology} presents our methodology in detail; Section \ref{sec:experiments} presents an empirical investigation on a test case in geophysical inversion; Section \ref{sec:conclusion} provides a conclusion and an outlook of future works.  

\section{Related works}
\label{sec:related_works}
Neural Networks (NN) have been used to estimate parameters of prescribed parametric families of distributions. \textit{Regression ABC} and \textit{Adaptive Gaussian Copula ABC} \cite{RegABCblum2010non,chen2019adaptivecopulaABC} train feed-forward network regressors $r(y)$ such that the posterior mean is approximated by $r(y_{obs})$, $y_{obs}$ being the observation vector. Initial samples are corrected such that the adjusted realizations correspond to a sample from the posterior. 
\textit{Sequential Neural Posterior Estimation} (SNPE) \cite[and references therein]{greenberg2019apt} sequentially approximates the posterior density where a NN takes $y_{obs}$ as input and outputs the parameters of a Gaussian mixture over the input space. 
\textit{Synthetic Neural Likelihood} (SNL) \cite{SNLpapamakarios2019} applies a similar iterative idea as SNPE but to approximate the likelihood by a Masked Autoregressive Flow network. Samples from the posterior are drawn by MCMC based on the synthetic likelihood. 
Neural density estimators were also used to learn proposal distributions integrated in MCMC \cite{kim2020sequential} or sequential Monte Carlo \cite{gu2015neural}.

\textit{Variational Auto-Encoders} (VAE) \cite{kingma2019VAEreview}, \textit{Generative Adversarial Networks} (GAN) \cite{creswell2018GANreview} and \textit{Invertible Neural Networks} (INN) \cite{kobyzev2020NFreview} avoid the explicit choice of parametric families for the distributions. Instead, they are trained to sample from distributions by transforming realizations from a simple (e.g. Gaussian or uniform) latent multivariate distribution into realizations from the distribution of interest by applying a sequence of non-linear transformations. In inversion, they were used to reduce the dimension of the input space in preparation for its exploration via the latent space, by MCMC \cite{mosser2020stochastic,LaloyLindeGAN2018,LaLoyLindeVAE2017} or by optimization \cite{richardson2018generative,lopez2020deep}.

In the previous methods the forward $F$ is called during training, inference or to sequentially guide the sampling from the input space. This is a limitation when $F$ is costly or only available through a sample of realizations. 

Learning direct transformations, parametrized by NN, from the observation vector into a plausible solution in the input space were investigated for inversion in geophysics \cite{laloy2019vec2pix,mosser2018rapid} and imaging \cite{mccann2017cnninverse,DLinverseimaging,ongie2020deepinverseimaging}. These approaches lack an inherent mechanism to quantify the variability in the proposed solutions. 
Conditional variants of generative NN, where a conditioning signal is provided along the latent random vector, were developed to directly sample from an approximation to the posterior distributions.  
cINN were demonstrated for inversion in astrophysics, medical imaging \cite{ardizzone2018cINN} and geophysics \cite{zabaras2020invertible}. 
These models impose a trade-off between tractable density estimation and sampling since most implementations have difficulties to compute inverses \cite{kobyzev2020NFreview}. A workaround is to train two networks: one for conditional sampling and one for density estimation \cite{zabaras2020invertible}. 
Pix2pix \cite{isola2017pix2pix} is a widely used cGAN \cite{mirza2014cGAN,ding2020ccgan} based framework for inversion in computer vision. cGANs were also demonstrated in medical image reconstruction \cite{adler2018deep} and in hydrogeology \cite{dagasan2020cGANhydro} (as a surrogate to the forward function).  
cVAE \cite{sohn2015cvae} were also used for similar purposes in computational imaging \cite{tonolini2020VAECI}. 
Conditional generative NN amortize inference by not calling the forward $F$ yet they lack an inherent mechanism to adapt to the unknown noise and to their approximation bias.

Our approach, in addition to amortizing inference by encoding the conditional distributions of interest by the jGNN, accounts for the noise without assumption on its distribution as well as for the jGNN training bias by sampling with ABC-SubSim and adapting its tolerance threshold. 
To the best of our knowledge, our work is the first to adapt joint generative modelling and ABC sampling on the latent space, based on Subset Simulation, to inverse problem solving. 
Among the surveyed methods from the literature, cVAE seems to be the closest to ours and we compare its performance to ours on the experimental test case. 

\section{Proposed methodology}
\label{sec:methodology}
We assume throughout the following that $(X,Y)$ can be expressed in function of some latent variable $Z$, of moderate dimension compared to $X$ and $Y$, such that $(X, Y) = G^{o}(Z) = (g_1^{o}(Z), g_2^{o}(Z))$. 
For a known $G^{o}$, uncovering the conditional distribution of $X$ knowing $Y=y$ amounts to uncovering the distribution of $g_1^{o}(Z)$ knowing $g_2^{o}(Z)$, which follows in turn from the conditional distribution of $Z$ knowing $g_2^{o}(Z)$. The goal of the following two sections is to present, first, the generative modelling framework used to estimate the map $G^{o}$, by $G$, from data and, second, how to approximate the conditional distribution of $Z$ knowing $Y=y$ by relying on ABC-SubSim. 

\subsection{Joint generative modelling}
The initial step is to train a joint Generative Neural Network (jGNN) to sample from the joint distribution of $(X,Y)$, denoted $P_{XY}$, based on available data $(X_1,Y_1),\dots,(X_n,Y_n)$. The jGNN is specified by a map $G: \mathbb{Z} \rightarrow \mathbb{X} \times \mathbb{Y}$ and a prior distribution $P_Z$ on $\mathbb{Z}$, where $\mathbb{X}$, $\mathbb{Y}$, $\mathbb{Z}$ are the domains in which $X,Y,Z$ vary, respectively. 

Training of the considered jGNN consists of minimizing some prescribed distance between $P_{XY}$ and $G_{\#}P_Z=P_{\Tilde{X} \Tilde{Y}}$, where $(\Tilde{X}, \Tilde{Y}) = (g_1(Z), g_2(Z))$ and $G_{\#}P_Z$ is the image (or \textit{pushforward}) probability measure of $P_Z$ by $G$. 

In our implementation we extended the Sinkhorn Auto-Encoder (SAE) \cite{patriniSinkhornAE}, a variant of the Wasserstein Auto-Encoder \cite{tolstikhin2018wassersteinae}, to the joint learning case by closely applying its formalism on an augmented space $\mathbb{X} \times \mathbb{Y}$.
Let $Q: \mathbb{X} \times \mathbb{Y} \rightarrow \mathbb{Z}$ represent the encoder map and $G: \mathbb{Z} \rightarrow \mathbb{X} \times \mathbb{Y}$ the decoder/generator map. 
The jGNN training goal is to minimize the optimal transport cost between $P_{XY}$ and $P_{\tilde{X}\tilde{Y}}$ via the minimization over deterministic maps $G$, in a family $\mathcal{G}$, of:
\noindent
\begin{equation*}
\begin{aligned}
W_c(P_{XY} ,P_{\tilde{X}\tilde{Y}}) &= \displaystyle\inf_{\pi \in \mathcal{P}(P_{XY} ,P_{\tilde{X}\tilde{Y}})} \mathbb{E}_{(X, \tilde{X}, Y, \tilde{Y}) \sim \pi} [c(X, \tilde{X}; Y, \tilde{Y})], \\
\end{aligned}
\label{eq:original_Wc}
\end{equation*}
where $\mathcal{P}(P_{XY} ,P_{\tilde{X}\tilde{Y}})$ is the set of all joint distributions having ``marginals'' (on $\mathbb{X}\times \mathbb{Y}$) $P_{XY}$ and $P_{\tilde{X}\tilde{Y}}$ and $c(.,.,.,.)$ is a function expressing the cost of transporting a couple $(X,Y)$ to a couple $(\tilde{X}, \tilde{Y})$. 
We consider $\mathbb{X}, \mathbb{Y}$ and $\mathbb{Z}$ to be Euclidean spaces and we set $c(X, \tilde{X}; Y, \tilde{Y}) = ||X-\tilde{X}||_p^p + ||Y-\tilde{Y}||_p^p$, a separable cost based on the $L_p$ norms on $\mathbb{X}$ and $\mathbb{Y}$. $c(.,.,.,.)$ is an $L_p$ norm, taken to the power $p$, on the product space $\mathbb{X} \times \mathbb{Y}$. 

Following \cite{patriniSinkhornAE}, the jGNN optimization objective is to minimize over deterministic maps $G$ and $Q$, in families $\mathcal{G}$ and $\mathcal{Q}$ respectively, of the quantity:
\begin{equation}
\begin{aligned}
        \mathcal{L} &= \sqrt[p]{\mathbb{E}_{XY \sim P_{XY}}[||X-g_1(Q(X,Y))||_p^p + ||Y-g_2(Q(X,Y))||_p^p]} + \lambda . W_p(Q_Z, P_Z),
\end{aligned}
\label{eq:jgnn_optobjective}
\end{equation}
with $\lambda$ to be greater than the Lipschitz constant of $G$. This loss balances between the objectives of reconstructing the training data accurately, while constraining the encoder $Q(Z|XY)$ to distribute its embeddings in $\mathbb{Z}$ such that $Q_{Z} = \mathbb{E}_{XY \sim P_{XY}}Q(Z|XY)$  fits a prescribed distribution $P_{Z}$. 

\begin{figure}[!tbh]
    \centering
    \captionsetup{justification=centering}
    \includegraphics[width = \textwidth]{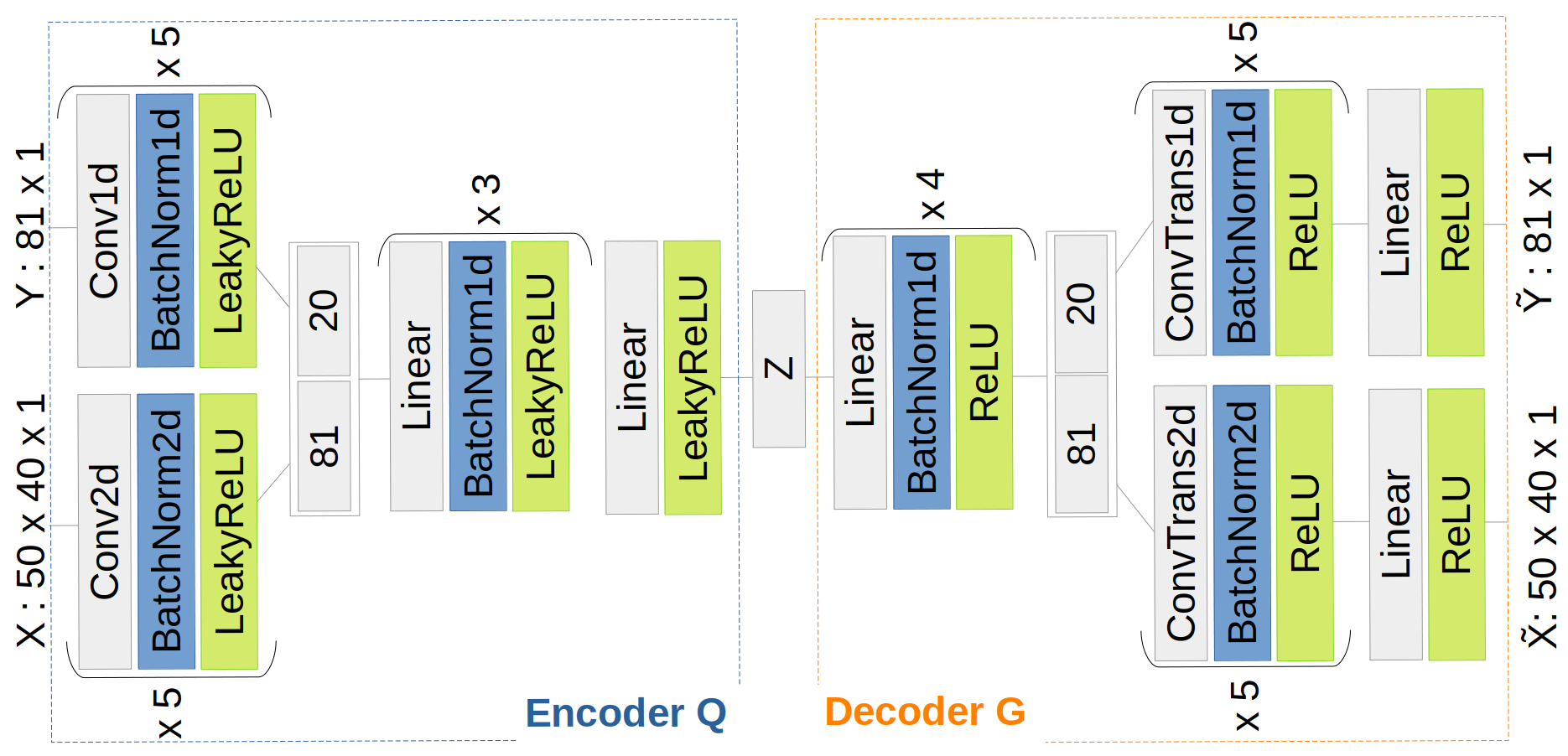}
    \caption{Schematic of the jGNN architecture and its components.}
    \label{fig:network_architecture}
\end{figure}

In practice, the $p$-th roots are removed from (\ref{eq:jgnn_optobjective}) for computational convenience. $W_p(Q_Z, P_Z)$ is estimated based on samples using the Sinkhorn algorithm \cite{Peyre2018GMsinkhorn}. To avoid the deterioration of the Wasserstein estimation when increasing the latent space dimension we set its entropy regularization parameter to 100 and its maximum number of iterations to 40. Optimization of the loss function was done with the Adam algorithm \cite{kingma2014adam} (lr = 0.001, $\beta_1$ = 0.9, $\beta_2$ = 0.999) with a batch size of 128. The reconstruction errors were taken as the $L_2$ norm normalized by the dimensions of $\mathbb{X}$ and $\mathbb{Y}$ (i.e. Mean Squared Errors). The parameter $\lambda$ was set to 150 at the beginning of the training and its value was cut by half every 500 epochs. This procedure seemed to help the training especially at the final stages where more weight is given to the reconstruction part of the loss compared to the regularization part. Spectral normalization \cite{miyato2018spectralnorm} was used in both the encoder and the decoder networks. In its absence we observed unstable training where one of the variables, $X$ or $Y$, is not learned correctly. The network architecture and its components are shown in Figure \ref{fig:network_architecture}. 

\subsection{Inversion by Subset Simulation}
We now consider that vector $y_{obs}$, assumed to be a realization from $Y$, was observed and we seek to retrieve the posterior distribution of $X$ knowing $Y=y_{obs}$. 
Given the deterministic jGNN outputting $\tilde{X} = g_1(Z)$ and $\tilde{Y} = g_2(Z)$, we can use instead the distribution of $g_1(Z)$ knowing $g_2(Z)=y_{obs}$. In other words, the posterior on $\mathbb{X}$ is induced by a posterior on $\mathbb{Z}$ and the jGNN surrogates the forward model during inversion. 
In practice, the equality $g_2(Z)=y_{obs}$ is seldom reached due to measurement errors on the observation vector and errors in the jGNN training, hence we need to consider instead that $y_{obs} = g_2(Z) + \eta$, where the term $\eta$ encompasses all sources of errors. To account for those errors, we introduce 
$\Gamma_{\epsilon} = \{z \in \mathbb{Z}: d(g_2(z)-y_{obs}) \leq \epsilon\}$ with $d(.,.)$ a dissimilarity measure on $\mathbb{Y}$ (e.g. its $L_p$ norm) and $\epsilon$ a tolerance parameter.
Furthermore, let $\pi_{Z}$ stand for the prior density of $Z$ (with respect to Lebesgue or some other dominating measure on $\mathbb{Z}$). We consider here a surrogate posterior density on $\mathbb{Z}$ (given $y_{obs}$) defined by $\pi_{Z|Z\in \Gamma_{\epsilon}} (z)\propto \mathbbm{1}_{\Gamma_{\epsilon}}(z)\pi_{\mathbb{Z}}(z)$. This posterior on $\mathbb{Z}$ directly leads to an approximate posterior distribution on $\mathbb{X}$ knowing $y_{obs}$ by image via $g_{1}$. 
Depending on the choice of $\epsilon$ and other problem settings such as the dimension of $\mathbb{Z}$, $\{Z \in \Gamma_{\epsilon}\}$ may become a rare event to simulate. We use SuS \cite{beck2001SS}, a rare event sampler, as an adaptive sampler in ABC \cite{chiachio2014approximate} to sample from $\pi_{Z|Z\in \Gamma_{\epsilon}}$. Our implementation of SuS is based on \cite{sus_acsCode}. 

SuS introduces a decreasing sequence of thresholds $+\infty = t_0 > t_1 > t_2 ... > t_m=  \epsilon$ 
which determines a sequence of nested subsets of  $\mathbb{Z}$, $\Gamma_{t_{\ell}} = \{z \in \mathbb{Z} : d(g_2(z),y_{obs}) \le t_\ell\} (\ell= 0, ..., m)$. 
For the sequence of events $\{Z \in \Gamma_{t_{\ell}}\}$ we have that : 
$P(Z \in \Gamma_{\epsilon}) =  P(Z \in \Gamma_{t_0}) \prod_{\ell=1}^m P(Z \in \Gamma_{t_{\ell}}|Z \in \Gamma_{t_{\ell-1}})$ with $P(Z \in \Gamma_{t_0})=1$ since $\{Z \in \Gamma_{t_0}\}$ is certain.  
This reduces the problem of estimating the small $p_{\epsilon}=P(Z \in \Gamma_{\epsilon})$ to estimating a sequence of larger conditional probabilities $P(Z \in \Gamma_{t_{\ell}}| Z \in \Gamma_{t_{\ell-1}})$. 

The SuS algorithm starts with an initial sample from the prior of $\mathbb{Z}$, $\{Z^{(0)}_i\}_{i=1}^N$, with a predefined size $N$. The dissimilarity values $\{d(g_2(Z^{(0)}_i), y_{obs})\}_{i=1}^N$ are calculated and ordered and the first threshold $t_1$ is defined as the $\alpha$-percentile of those values. $\alpha$ is prescribed and typically chosen in the range $[0.1, 0.3]$ \cite{zuev2012bayesian}. The set $\Gamma_{t_1}$ is first populated by the observations from this initial sample that yield distances below $t_1$. Starting from each one of those succeeding observations, sufficiently many states of a Markov chain with stationary distribution $\pi_{Z|Z\in \Gamma_{t_1}}$ are generated to complete the current elements of $\Gamma_{t_1}$ up to $N$ elements (cf. \cite{papaioannou2015mcmc} for specific details on the MCMC sampling methods with SuS). At each subsequent iteration $\ell = 2, ..., m$, the sample $\{Z^{(\ell-1)}_i\}_{i=1}^N$ is used to calculate $\{d(g_2(Z^{(\ell-1)}_i), y_{obs})\}_{i=1}^N$ and to set $t_{\ell}$ as the $\alpha$-percentile of those distances. New observations in $\Gamma_{t_{\ell}}$ are again sampled starting from the observations that yield distances below $t_{\ell}$.  This process stops when $\epsilon$ is crossed (i.e. if the proposed $t_{\ell} \leq \epsilon$ then $m$ is defined as $\ell$ and $t_{m}$ is set equal to $\epsilon$) or when a prescribed maximum number of iterations is reached.
The final elements of $\Gamma_{\epsilon}$ are used to form candidate solutions in $\mathbb{X}$ via $g_1$.
$p_{\epsilon}$ can be estimated via $\widehat{p_{\epsilon}}=\alpha^{m-1}\frac{N_{m-1}}{N}$, with $N_{m-1}$ being the number of succeeding particles at the penultimate iteration of SuS. This estimator sheds some light on the diversity/uncertainty in the proposed solutions, from the jGNN's latent space perspective.

\section{Experiments}
\label{sec:experiments}
\begin{figure}[!htb]
    \centering
    \includegraphics[width =\textwidth]{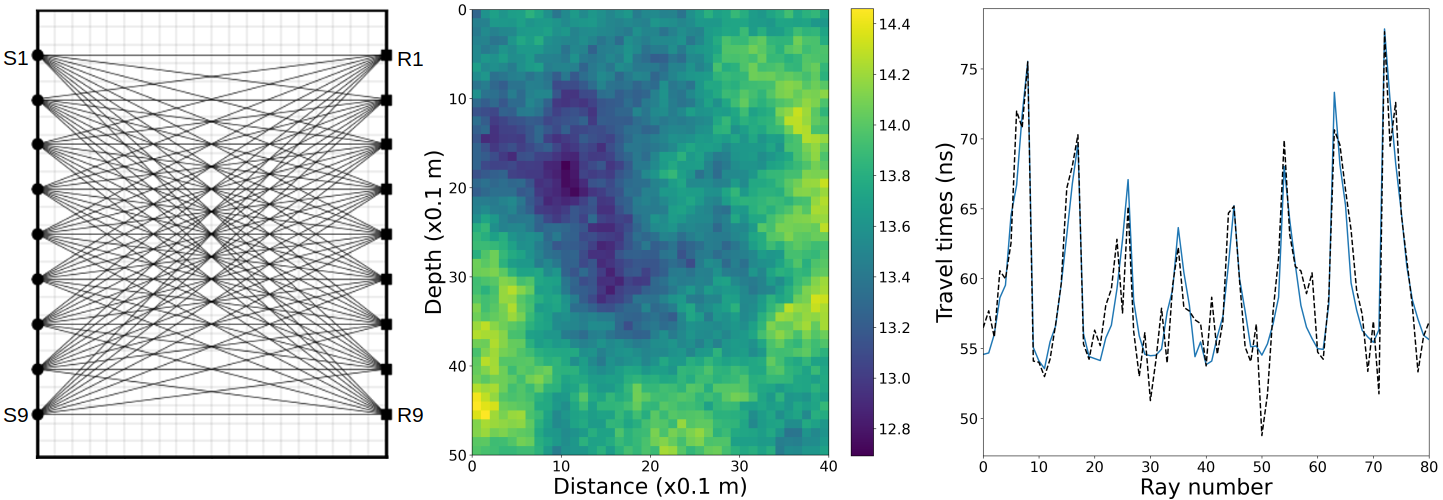}
    \caption{(a): Cross-hole tomography setup, S1-9 GPR sources and R1-9 GPR receivers; (b): slowness field in ns/m; (c): first arrival travel time vectors in ns given by the forward solver corresponding to (b) and contaminated by standard Gaussian noise realizations with levels of 0.54 ns (blue-solid) and 2.29 ns (black-dashed).}
    \label{fig:models_example}
\end{figure}

To showcase the methodology in realistic inversion settings, we applied it to cross-hole Ground Penetrating Radar (GPR) tomography. In this geophysical method, a source emits high-frequency electromagnetic waves at a given depth in one borehole, while the response is recorded by a receiver antenna at a given depth in an adjacent borehole. The first-arrival travel times of the recorded traces, for different acquisition geometries, are used to retrieve the slowness field (i.e. inverse of the velocity field) between the boreholes.\\
Training and test sets of couples of subsurface domains and their corresponding solver output (i.e. $(X,Y)$) were simulated using an approximate linear forward solver (Figure \ref{fig:models_example}(b,c)).  
The domain (i.e. $X$) is discretized on a grid of size 50 $\times$ 40 with a cell size of 0.1 m, leading to $\mathbb{X}$ being of dimension 2000. The boreholes are located 3.9 m apart (Figure \ref{fig:models_example}(a)). Nine source and receiver locations are regularly spaced between 0.5 and 4.5 m depth leading to a measurement vector (i.e. $Y$) with 81 travel times. The slowness field is described by a Gaussian prior with an isotropic exponential kernel, a length scale of 2.5 m and a variance at the origin of 0.16 (ns/m)$^2$. 

\begin{figure}[!tbh]
    \centering
    \includegraphics[width = \textwidth]{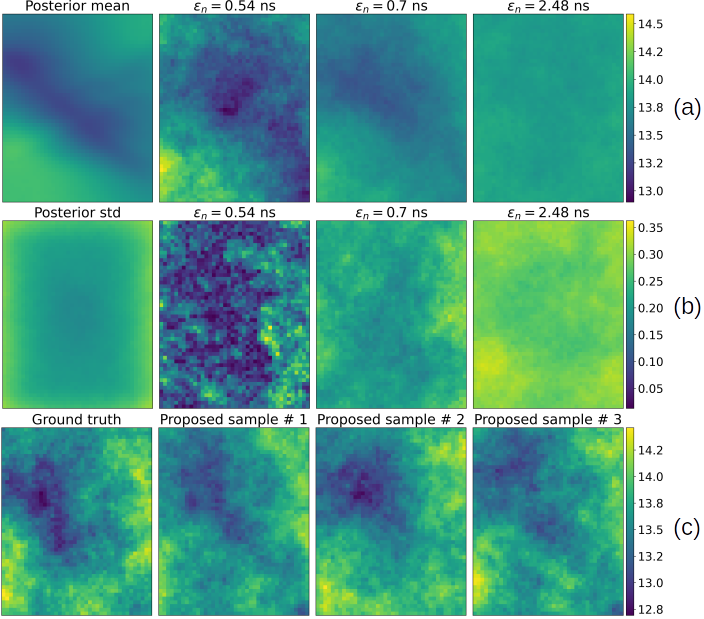}
    \caption{Example showing:(a) pixel wise means of the exact Gaussian posterior and proposed samples by our method at different thresholds $\epsilon_n$; (b) pixel wise standard deviations of the exact Gaussian posterior and proposed samples by our method at different thresholds $\epsilon_n$; (c) ground truth slowness field along proposed solutions by our method at $\epsilon_n$ = 0.7 ns. Gaussian noise realization with a standard deviation of 0.54 ns.}
    \label{fig:means_std_samples}
\end{figure}

\begin{figure}[!htb]
 \centering
    \includegraphics[width=\textwidth]{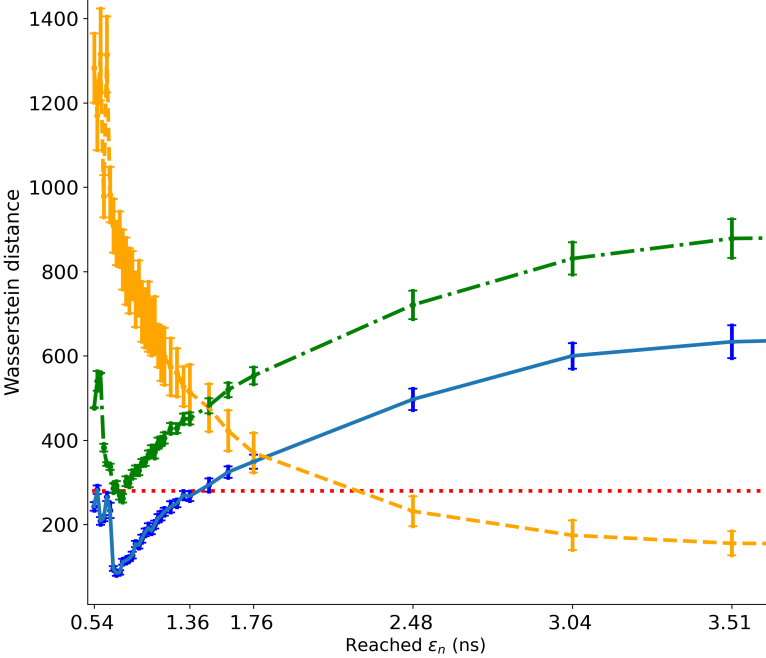}
    \caption{Example of sample-based estimates of the Wasserstein distances between: (blue-solid) our approximate posterior and the exact Gaussian posterior; (orange-dashed) our approximate posterior and the prior on $\mathbb{X}$; (green-dot-dash) our approximate posterior and the Dirac located at the true solution. The red dotted line is a sample-based estimate of the Wasserstein distance between the exact posterior and the Dirac distribution located on the true solution. The minima of the green-dot-dashed and the blue-solid are located in the interval [0.66 ns, 0.74 ns] of $\epsilon_n$. Gaussian noise realization with a standard deviation 0.54 ns.}
    \label{fig:wass_distance}
\end{figure}

The data used for training were not noise-contaminated. To validate the method, we sample solver outputs from the test set, not seen during training, and contaminate them with noise vectors and use the noisy vectors as the measurement vectors, $y_{obs}$, to invert. In the following, we present results obtained for noise vectors from the standard multivariate Gaussian with a standard deviation of 0.5 ns (referred to by "small noise") and a standard deviation of 2.5 ns (referred to by "large noise"). Furthermore, working under the unknown noise assumption, we did not contaminate the jGNN proposed travel times ($\tilde{y}$) by noise during the ABC-SubSim posterior approximation, as is done in classical ABC. 
This example allows to manipulate moderately high dimensional variables while still providing analytical solutions as basis for comparison. Namely, when the noise is Gaussian, given the Gaussian prior on the field and the linear solver, the exact Gaussian posterior on $\mathbb{X}$ is available analytically \cite{tarantola2005inverse}. We present the results on forty inversions with different slowness fields and contaminated with different noise realizations. The metric used for comparison is the Root Mean Squared Error (RMSE)\footnote{For vectors $V_1$ and $V_2$, both of dimension $m$, $RMSE(V_1,V_2) = \sqrt{\frac{1}{m}||V_1-V_2||_2^2}$}.
\paragraph{ABC-SubSim threshold impact:} in practice, we ran the ABC-SubSim posterior approximation to retrieve solutions that guarantee $||\tilde{y}-y_{obs}||_2^2 \le \epsilon$ with targeted threshold $\epsilon \in$ [0.01 ns$^2$, 3000 ns$^2$]. In the following $\epsilon_n$ refers to $\sqrt{\frac{\epsilon}{81}}$ ns, the normalized value of $\epsilon$. 
When the targeted value for $\epsilon$ is close or below the noise level, the SuS algorithm consumes its iterations budget before reaching the $\epsilon$ by the sequential update loop and stagnates at threshold values close to the noise level (its $L_2$ squared norm). At these thresholds, ABC-SubSim is trying to fit the noise very closely which leads to low diversity solutions and potential artifacts (Figure \ref{fig:means_std_samples}(a,b) column ``$\epsilon$= 0.54 ns''). On the contrary, when the threshold is very large, the proposed solutions move further away from the ground truth and the posterior approximation moves closer to the prior distribution (Figure \ref{fig:means_std_samples}(a,b) column ``$\epsilon$= 2.48 ns''). Between these extremes lies a set of values for the threshold that provide an approximate posterior whose samples show comparable statistics to the analytical posterior samples (Figure \ref{fig:means_std_samples}(a,b) column ``$\epsilon$= 0.7 ns''). This set can be seen on Figure \ref{fig:wass_distance} where the sample-based estimates of the Wasserstein distance between our approximate posterior and, on one hand, the ground truth Dirac, and the exact posterior on the other hand, are minimal.  

\begin{figure}[!htb]
    \centering
    \includegraphics[width = \textwidth]{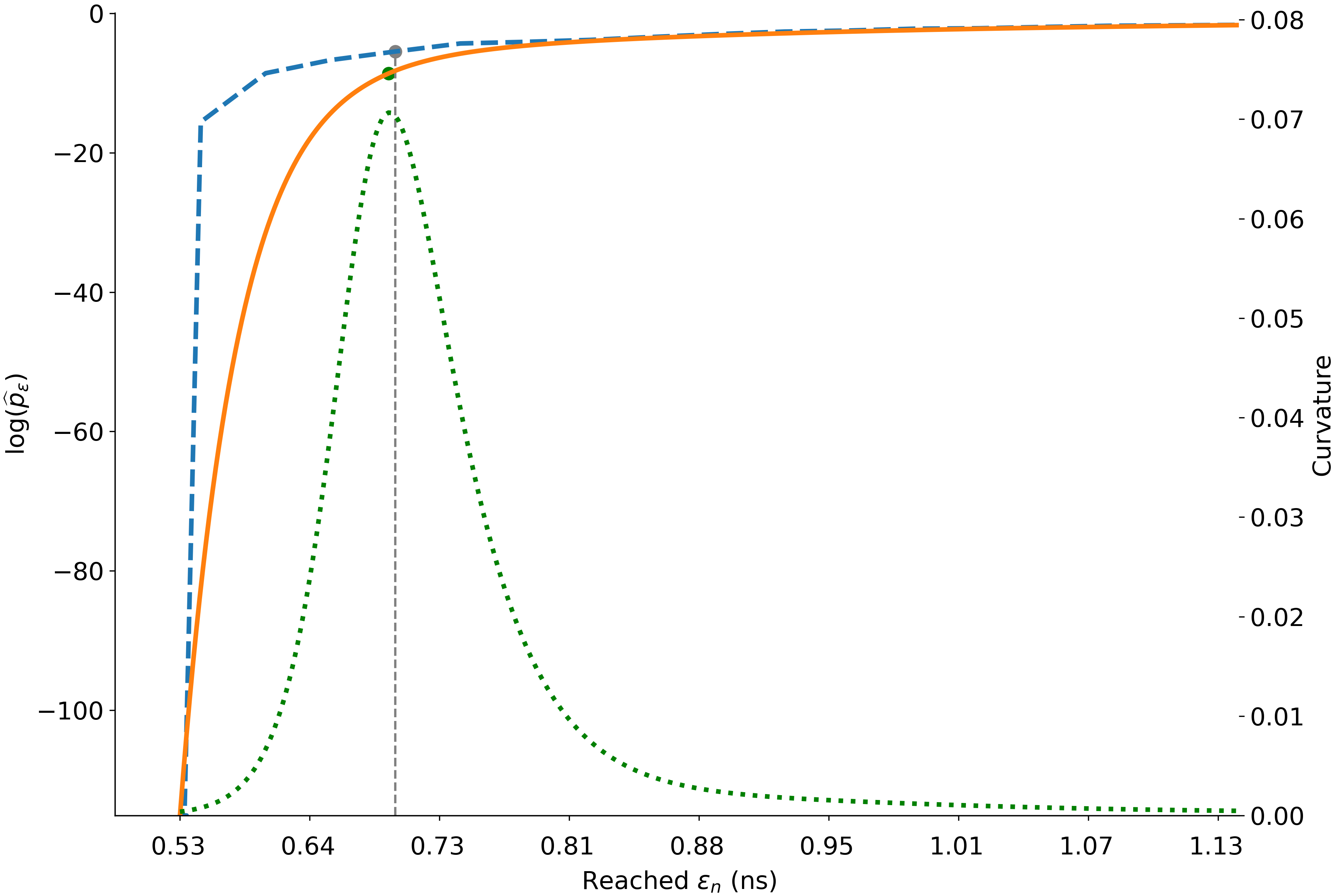}
    \caption{Example of (blue-dashed) logarithm of the estimate $\widehat{p_{\epsilon}}$ of the probability that a randomly sampled $z$ belongs to the solution set $\Gamma_{\epsilon}$ and (orange-full) its smoothed version used to estimate the (green-dotted) curvature. The peak of the curvature suggests an adequate threshold for the ABC-SubSim approximation. We take the closest threshold in the set of tested threshold values (grey-dashed), in this case $\epsilon_n$ = 0.7 ns. Gaussian noise realization with a standard deviation 0.54 ns.}
    \label{fig:pf_curvature}
\end{figure}

However, the plots in Figure \ref{fig:wass_distance} are not available in practice and cannot be used to select a suitable threshold for ABC-SubSim. Instead, we propose to monitor the evolution of $\widehat{p_{\epsilon}}$, an estimate of the probability that a randomly sampled $z$ belongs to the solution set $\Gamma_{\epsilon}$ and provided by the SuS algorithm. The probability content curve in Figure \ref{fig:pf_curvature} provides two important pieces of information: the lowest value reached by $\epsilon_n$ on the horizontal axis, where the sequential update to the ABC-SubSim threshold by SuS stagnates, is informative about the (normalized) noise level; and the values of $\epsilon_n$ falling at, or in close proximity to, the point of highest curvature coincide with the region of the most suitable thresholds identified in Figure \ref{fig:wass_distance}. The curvature is estimated based on a smoothed version of the logarithm of the probability $\widehat{p_{\epsilon}}$ as function of $\epsilon$.

\begin{figure}[!htb]
    \centering
    \includegraphics[width = \textwidth]{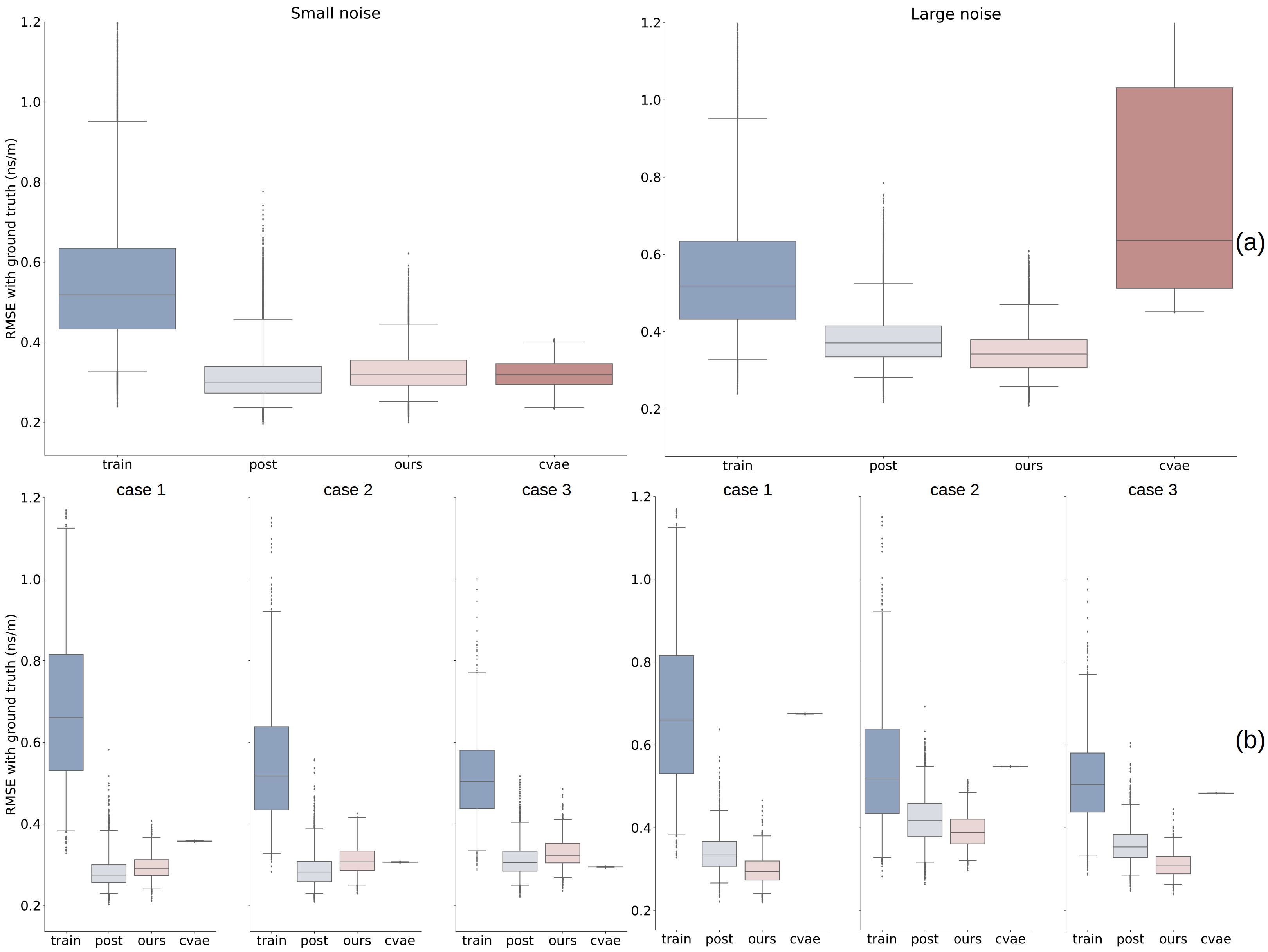}
    \caption{(a) Distributions of the aggregated RMSE values between proposed solutions and ground truths for all forty inversion tests, with two noise scenarios: (left column) realizations from a standard Gaussian with standard deviation of 0.5 ns; (right column) realizations from a standard Gaussian with standard deviation of 2.5 ns. On the horizontal: "train" refers to RMSE values between ground truths and training data set; "post" refers to RMSE values between analytical posterior samples and ground truths; "ours" refers to RMSE values between our ABC-SubSim proposed solutions and ground truths at the threshold at maximum curvature of $\log(\widehat{p_{\epsilon}})$; "cvae" refers to RMSE values between our trained cVAE proposed solutions and ground truths. These plots are at latent space size of 30 and training set size of 1000. The plots are cut off above RMSE values of 1.2 ns to reduce space. (b) Three specific inversion cases disposed as in (a) with regards to the noise and the horizontal axis.}
    \label{fig:allinv_big_smallnoise_cvae}
\end{figure}

\paragraph{Comparison with cVAE:} we compared our methodology with a conditional distribution sampler learned by a cVAE. For space limitation, we refer the reader to \cite{sohn2015cvae,kingma2019VAEreview,kruse2021benchmarkingINNinverse,ren2020benchmarking} for formal and implementation details. In summary, the main differences between our jGNN and a cVAE are: the jGNN is required to reconstruct both $X$ and $Y$ while cVAE only needs to reconstruct $X$ given a realization of $Y$; and the encoder in cVAE aims to fit $Q(Z|X,Y)$ with a standard Gaussian for each couple $(X,Y)$ while our jGNN aims to fit the aggregated $Q(Z)$ for all couples $(X,Y)$ with the standard Gaussian. \\  
We adapted our jGNN to the cVAE training objective in order to have models with comparable number of trainable parameters. Both were trained for 5000 epochs with the Adam optimizer (learning rate = 0.001, $\beta_1$ = 0.9, $\beta_2$ = 0.999). We monitored reconstruction statistics on a validation set and picked the best training epoch towards the end of the training. To select the regularization parameter for cVAE we tested several values in \{0.005, 0.05, 0.585, 1\} and kept 0.05 which gave the best reconstruction statistics on the validation set. Under these conditions, we see in Figure \ref{fig:allinv_big_smallnoise_cvae}((a)-left column) that overall the cVAE model is able to retrieve solutions that are close to the ground truth when the noise is small. However, when looking at individual inversions results in Figure \ref{fig:allinv_big_smallnoise_cvae}(b) we see that the diversity of those solutions is very low compared to the analytical posterior. This is an indicator of the posterior collapsing to a single mode, a behaviour similarly observed with cVAE in other contexts \cite{kruse2021benchmarkingINNinverse,razavi2018postcollapse}. Furthermore, Figure \ref{fig:allinv_big_smallnoise_cvae} ((a)-right column) shows that the cVAE model we trained is unable to retrieve meaningful solutions in the large noise situation while our methodology remains robust in such conditions at a suitable ABC-SubSim threshold.

\paragraph{Assessment of the learned forward function:} as in \cite{kruse2021benchmarkingINNinverse}, we refer by ``resimulation'' to the output of the forward function $F$ called on our proposed solutions $\tilde{X}$ and we denote them by $Y_r$ (i.e. $Y_r = F(\tilde{X})$). 
RMSE between $Y_r$ and the output of the jGNN $\tilde{Y}$, generated along the proposed solutions $\tilde{X}$, estimate the errors of the learned function $F$ by the jGNN. For ``small noise'' realizations, these RMSE values had an average of 0.370 ns (median: 0.346 ns; 95\% interval: [0.194 ns, 0.687 ns]) and an average of 0.424 ns (median: 0.385 ns; 95\% interval: [0.204 ns, 0.834 ns]) for the ``large noise'' realizations. Even though no common reference for comparison is available, these values should tend to zero as the jGNN training is improved. 

Furthermore, the RMSE values calculated between $Y_r$ and $y_{obs}$ were on average at 0.745 ns (median: 0.729 ns; 95\% interval: [0.571 ns, 1.015 ns]) with the ``small noise'' realizations and at 2.500 ns (median: 2.489 ns; 95\% interval: [2.206 ns, 2.888 ns]) with ``large noise'' realizations. These values should be close to the level of the noise that contaminated the observations which is often the case with the ``large noise'' scenario (i.e. noise standard deviation of 2.5 ns). In the ``small noise'' scenario we believe that the jGNN bias is having a more perceivable impact on the sampling of potential solutions explaining the discrepancy with the noise (i.e. noise standard deviation of 0.5 ns). 

\paragraph{Latent space dimension and training set size impact:} we trained the jGNN with dimensions of the latent space in \{10, 30, 100\} and with training set sizes in \{1000, 4000\}. For lack of space we do not detail these results here. Yet, in summary, very comparable performances across dimensions were observed for the training set size of 1000. This suggests that the inversion with ABC-SubSim is robust with regards to the latent space dimension choice. Similar observations were made with training set size of 4000 with slight deterioration for dimension 10 and 30, hinting to a potential overfit at these dimensions.

\section{Conclusion and outlook} \label{sec:conclusion}
We proposed a methodology for realistic inverse problems based on learning a joint deep generative model constructing a low dimensional latent space encoding the variables of the problem. On this latent space an approximate posterior is sampled from by ABC with the Subset Simulation algorithm locating adequate regions for the solutions. Our experiments show promising potential when compared to the analytical solution and a conditional VAE. Furthermore, the method demonstrates robustness to large levels of noise, often encountered in practice. Robustness was also observed across the latent space dimension choice. Furthermore, we proposed to monitor the evolution of the probability content in the latent space. From the resulting diagnostic curve, we retrieved an indication about the unknown noise level and we identified an empirical rule to set the ABC-SubSim threshold based on locating its maximum curvature. Finally, in our framework calling the forward solver was avoided during inference which is expected to keep the computational cost of the method relatively low across multiple inversions, after the initial training data set generation. 

To improve on this work, we will extend the evaluation to inverse problems with non-linear physics and different noise distributions. Similarly, analysing the sensitivity of the approach to the norms choice in the jGNN objective and ABC-SubSim, the prior distributions of $Z$ and $X$, and the network architecture is paramount to further validate the methodology. Finally, augmenting our approach with an adaptive scheme to optimally sample new training data points and integrating them on the fly can help improve its sample efficiency.   

\clearpage

\bibliographystyle{abbrv}

\end{document}